\documentclass[oneside,english]{amsart}
\usepackage[T1]{fontenc}
\usepackage[latin9]{inputenc}
\usepackage{units}
\usepackage{amsthm}
\usepackage{amssymb}
\usepackage{amsmath}
\usepackage{hyperref}
\usepackage{listings}

\usepackage{tikz}
\usetikzlibrary{shapes,snakes}
\usetikzlibrary{decorations.markings}
\usetikzlibrary{backgrounds}

\makeatletter
        \newtheorem{theorem}{Theorem}[section]
       \theoremstyle{definition}
          \newtheorem{definition}[theorem]{Definition}
    \newtheorem{example}[theorem]{Example}
    \newtheorem{lemma}[theorem]{Lemma}

\makeatother

\usepackage{babel}

\newcommand{\code}{\texttt}
\begin{document}

\title{Technical Report \\ The Stochastic Network Calculator}

\author{Michael Beck, Sebastian Henningsen}

\maketitle

\begin{center}
\email{michaelalexbeck@gmail.com, sebastian.henningsen@hu-berlin.de}

\date{February, 2017}
\end{center}

\section{Introduction and Overview}

In this technical report, we provide an in-depth description of the Stochastic Network Calculator tool, henceforth just called ``Calculator''. This tool is designed to compute and automatically optimize performance bounds in queueing networks using the methodology of stochastic network calculus (SNC).
For a detailed introduction to SNC, see the publicly available thesis of Beck \cite{Beck:thesis}, which shares the notations and definitions used in this document. Other introductory texts are the books \cite{Chang:book, Jiang:book} and the surveys \cite{Fidler:survey, Fidler:guide}.

The Stochastic Network Calculator is an open source project and can be found on github at \url{https://github.com/scriptkitty/SNC}; it was also presented in \cite{Beck:SNCalc, Beck:SNCalc2}.
We structure this report as follows:
\begin{itemize}
    \item We give the essential notations, definitions, and results of SNC in Sections \ref{sec:Essentials} and \ref{sec:Theoretical Results}. Section \ref{sec:Modeling with SNC} focuses on modeling queueing systems, service elements, and arrivals within the language of SNC.
    \item Section \ref{sec:Code Structure} gives an overview on the calculator's code structure and its workflow.
    \item In Section \ref{sec:Code Representation} we give more detail on how the concepts of SNC are represented in the code. 
    \item How to access and extend the Calculator is topic of Section \ref{sec:APIs and Extensions}.
    \item We wrap things up with a full example in Section \ref{sec:full_example}. It describes the modeling steps we have undertaken to produce the results for our presentation at the IEEE Infocom conference 2017 \cite{Beck:SNCalc2}.
\end{itemize}

We refrain from displaying larger chunks of code in this report for two reasons: (1) The Calculator is under development and the code-base might change at any time. (2) The code is extensively commented; hence, instead of enlargening this report to unbearable lengths, it will be more useful for developers to read about the code's details in their own context.

\section{Essentials of Stochastic Network Calculus}\label{sec:Essentials}

This Section orients itself on the notations and definitions made in \cite{Beck:thesis}.

We partition time into time-slots of unit length and consider a fluid data model. In this scenario we define a \emph{flow} as a cumulative function of data:

\begin{definition}
A \emph{flow} is a cumulative function 
\begin{align*}
    A\,:\,\mathbb{N}&\rightarrow \mathbb{R}_0^+ \\
    t&\mapsto A(t)
\end{align*}
\end{definition}
The interpretation of \(A(t)\) is the (cumulative) amount of data arriving up to time \(t\). Correspondingly the doubly-indexed function \(A(s,t)\) describes the amount of data arriving in the interval \((s,t]\). 

In SNC a stochastic bound on the amount of arrivals is needed. Without such a bound the total number of arrivals in some interval could be arbitrarily large, thus, making an analysis of the system impossible. The Calculator is based on the MGF approach of SNC. Flows and other stochastic processes are represented by their respective MGF which are upper bounded.

\begin{definition}\label{def:Arrival-Bound}
The MGF of some quantity \(A(t)\) at \(\theta\) is defined by
\[\phi_{A(t)}(\theta) = \mathbb{E}(e^{\theta A(t)}),\]
where \(\mathbb{E}\) denotes the expectation of a random variable.

We have an MGF-bound for a flow \(A\) and the value \(\theta>0\), if
\[\phi_{A(s,t)}(\theta) \leq e^{\theta \rho(\theta)(t.s) + \theta \sigma(\theta)}\]
holds for all time pairs \(s\leq t\).
\end{definition}
\begin{figure}
    \centering
\begin{tikzpicture}[scale = 1]   

\path (0,0) node(start) {} 
	++(1,0) node[circle,draw](U) {$U$} 
	++(1,0) node(end) {}; 
\draw[->] (start) -- node[above] {$A$}(U);
\draw[->] (U) -- node[above] {$B$}(end);

\end{tikzpicture}
    \caption{A single flow \(A\) enters a single service element. Its output is denoted by flow \(B\).}
    \label{fig:single-queue}
\end{figure}
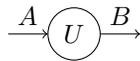
The second basic quantity in a queueing system is the amount of service per time-slot. The relationship between these two and the resulting output (see\break Figure \ref{fig:single-queue}) is given by
\[B(t) \geq A\otimes U(0,t) = \min_{0\leq s\leq t}\{A(0,s)+U(s,t)\}.\]
Here \(U\) is a bivariate function (or stochastic process) that describes the service process's behavior. For example a constant rate server takes the form\break \(U(s,t) = r_U(t-s)\), where \(r_U\) is the service element's rate.

\begin{definition}\label{def:Service-Bound}
A service element is a dynamic \(U\)-server, if it fulfills for any input-output pair \(A\) and \(B\):
\[B(t)\geq A\otimes U(0,t).\]

Such a server is MGF-bounded for some \(\theta>0\), if \(U\) fulfills
\[\phi_{U(s,t)}(-\theta)\leq e^{\theta\rho(\theta)(t-s) + \theta\sigma(\theta)}\]
for all pairs \(s\leq t\).
\end{definition}

Note the minus-sign in the above definition to indicate that the service is bounded from below, whereas the arrivals are bounded from above.

We are particularly interested in two performance measures that puts a system's arrivals and departures into context.

\begin{definition}
The \emph{backlog} at time \(t\) for an input-output pair \(A\) and \(B\) is defined by
\[\mathfrak{b}(t) = A(t) - B(t).\]
The \emph{virtual delay} at time \(t\) is defined by
\[\mathfrak{d}(t) = \min\{s\geq 0 \mid A(t) \leq B(t+s)\}.\]
\end{definition}

Note that in these definitions we make two assumptions about the queueing system: (1) As the backlog is defined by the difference of \(A\) and \(B\), we assume the system to be loss-free -- all data that has not yet departed from the system must still be queued in it. (2) We only consider the \emph{virtual} delay. This is the time until we see an amount of departures from the system, which is \emph{equivalent} to the accumulated arrivals up to a time \(t\). For FIFO-systems its virtual delay coincides with its delay; in non-FIFO systems, however, this does not need to be the case.

\section{Modeling with Stochastic Network Calculus}\label{sec:Modeling with SNC}

\subsection{Modeling the Network}

\begin{figure}
    \centering
    \begin{tikzpicture}[scale = 1]   

\path (-1,1) node(source_1)[align=center] {Source}
	  (0,-1) node(source_2)[align=center] {Source};

\draw[->] (source_1) -- (0,1);
\draw[->] (source_2) -- (1,-1);

\draw (0,0.75) -- ++(0.75,0) -- ++(0,0.5) -- ++(-0.75,0);
\draw (1,-1.25) -- ++(0.75,0) -- ++(0,0.5) -- ++(-0.75,0);

\draw (0.65,0.8) -- ++(0,0.4);
\draw (0.6,0.8) -- ++(0,0.4);

\draw (1.65,-1.2) -- ++(0,0.4);
\draw (1.6,-1.2) -- ++(0,0.4);
\draw (1.55,-1.2) -- ++(0,0.4);

\path (1,1) node[circle, draw, minimum size = 0.5cm](Server_1){}
	  (2,-1) node[circle, draw, minimum size = 0.5cm](Server_2){};

\draw[->] (Server_1) -- (3.5,0.3);
\draw[->] (Server_2) -- (3.5,-0.3);

\draw (3.5,0.05) -- ++(0.75,0) -- ++(0,0.5) -- ++(-0.75,0);
\draw (3.5,-0.55) -- ++(0.75,0) -- ++(0,0.5) -- ++(-0.75,0);

\draw (4.15, 0.1) -- ++(0,0.4);
\draw (4.1, 0.1) -- ++(0,0.4);
\draw (4.05, 0.1) -- ++(0,0.4);

\draw (4.15, -0.5) -- ++(0,0.4);
\draw (4.1, -0.5) -- ++(0,0.4);

\path (4.85,0) node[circle, draw, minimum size = 1.2cm](Server_3){}
	++(0,0.3) node[circle, minimum size = 1.2cm](dummy_top){}
	++(0,-0.6) node[circle, minimum size = 1.2cm](dummy_bottom){};

\path (7,0.3) node[align = center](Destination_1){Departures}
	  (7,-0.3) node[align = center](Destination_2){Departures};

\draw[->] (dummy_top) -- (Destination_1);
\draw[->] (dummy_bottom) -- (Destination_2);

\end{tikzpicture}\break
(a)\break
\begin{tikzpicture}[scale = 1]   

\path (-0.75,0) node(source)[draw, circle] {$e$}
	  (1,1) node(U)[draw, circle] {$U$}
	  (1,-1) node(V)[draw, circle] {$V$}
	  (3,0) node(W)[draw, circle] {$W$}
	  ++(0.05,0.25) node(top_dummy)[circle, text = white] {$\cdot$}
	  ++(0,-0.5) node(bottom_dummy)[circle, text = white] {$\cdot$}
	  ++(2,0.25) node(destination)[draw, circle] {$e^\prime$}
	  ++(-0.2,0.05) node(dest_top_dummy)[circle, text = white] {}
	  ++(0,-0.1) node(dest_bottom_dummy)[circle, text = white] {};

\draw[->] (source) -- (U);
\draw[->] (U) -- (top_dummy);
\draw[->] (top_dummy) -- (dest_top_dummy);
\draw[->] (source) -- (V);
\draw[->] (V) -- (bottom_dummy);
\draw[->] (bottom_dummy) -- (dest_bottom_dummy);

\end{tikzpicture}\break
(b)\break
    \caption{The original queueing system (a) and its graph representation (b). The nodes with labels \(e\) and \(e^\prime\) are the ``outside'' of the network.}
    \label{fig:graph-representation}
\end{figure}
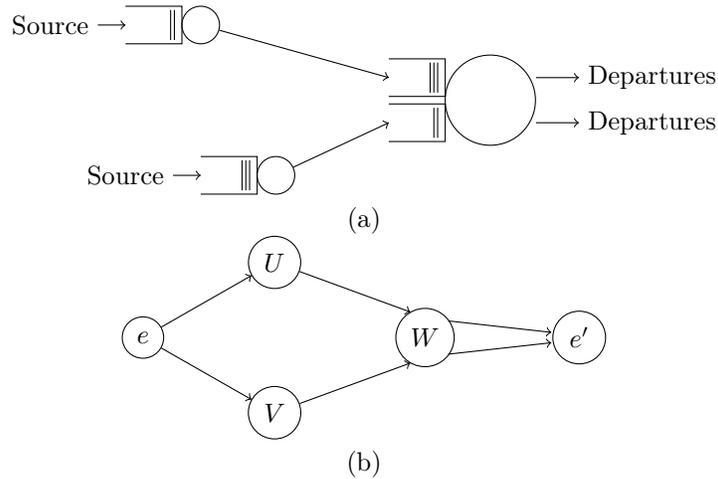
In SNC we consider a queueing system such as a communication network as a collection of flows and service elements. These can be represented as nodes and edges as shown in Figure \ref{fig:graph-representation}. In this transformation we replace each flow's route by a sequence of directed edges, such that each hop of the flow through the network is mapped to one edge; furthermore, we  introduce two extra nodes. They represent the ``outside'' of the network. Each flow originates from \(e\) and leaves the network via \(e^\prime\).
Note that in such a scenario there does not need to be a one to one correspondence between nodes and physical entities. In the graph representation one node with several inputs just represents one resource that is expended by several flows of arrivals.

For example: A router can have several interfaces each leading to another router. In this scenario data packets do not queue up in the router, but rather in each of its interfaces; hence, the nodes in the corresponding graph represent a single interface only and not the entire router. This leads to different topologies between the physical network (in terms of routers and their connections) and the graph of service elements and flows.

\subsection{Modeling the Arrival Processes}

Now, we give more details on the modeling of data flows and present some of the arrival bounds currently implemented in the tool. As introduced in the previous chapter we use a fluid model with discrete time-slots in the Calculator. This means we are interested in the arrival's distribution per time-slot; or more precisely: in their moment generating function (MGF). The easiest (and bland) example is a stream of data with constant rate:

\begin{example}
Assume a source sends \(r\) data per time-slot (for example\break 100 Mb/sec.). It immediately follows: \(A(s,t)=r(t-s)\) and as there is no randomness involved it also holds \(\phi_{A(s,t)}(\theta)=e^{\theta r(t-s)}\) for the MGF. To achieve a bound conform to Definition \ref{def:Arrival-Bound} we define \(\rho(\theta)=r\) and \(\sigma(\theta)=0\).
\end{example}

We can construct a simple random model by assuming that the arrivals per time-slot are stochastically independent and follow the same distribution (an i.i.d. assumption).
\begin{example}\label{ex:Exponential-Increments}
Assume a source sends in time-slot \(t\) an amount of data equal to \(a_t\). Here \(a_t\) are stochastically independent and exponentially distributed with a common parameter \(\lambda\). The exponential distribution has density
\[f(x) = \lambda e^{-\lambda x} \qquad \text{ for all } x\geq 0\]
and \(f(x)=0\) for all x < 0. From this we can derive the MGF for a single increment as \(\phi_{a(t)}(\theta)=\frac{\lambda}{\lambda-\theta}\). Due to the stochastic independence of the increments we have:
\[\phi_{A(s,t)}(\theta)=\prod_{r=s+1}^t \phi_{a(r)}(\theta)=\left(\frac{\lambda}{\lambda - \theta}\right);\]
hence, an MGF-bound for this type of arrivals is given by \(\rho(\theta) = 1/\theta \log(\frac{\lambda}{\lambda - \theta})\) and \(\sigma(\theta) = 0 \).
\end{example}
The above example can be easily generalized to increments with other distributions, as long as their MGF can be computed.

As of today we can roughly divide the methodology of SNC by the way of how to bound the involved stochastic processes (for details refer to \cite{Fidler:survey}). The Calculator uses MGF-bounds as in Definitions \ref{def:Arrival-Bound} and \ref{def:Service-Bound}. In the next example we show how to convert bounds from the ``other'' branch of SNC to MGF-bounds.

\begin{example}
We say a flow has exponentially bounded burstiness (follows the EBB-model), if 
\[\mathbb{P}(A(s,t)>\rho(t-s)+\varepsilon)\leq Me^{d \varepsilon}\]
holds for all pairs \(s\leq t\) and \(\varepsilon>0\). In this model we call \(M\) the prefactor and \(d\) the bound's decay; the parameter \(\rho\) represents the arrival's long-term rate.
We can convert such a bound to an MGF-bound (see for example \cite{Li_eff_bandwidth_2007} or Lemma 10.1 in \cite{Beck:thesis}) via
\[\phi_{A(s,t)}(\theta) \leq \int_0^1 e^{\theta (\rho(t-s)+\varepsilon)}\mathrm{d}\varepsilon^\prime.\]
Here \(\varepsilon = -1/d \log(\tfrac{\varepsilon^\prime}{M})\). Solving the above integral leads to
\[\phi_{A(s,t)}(\theta) \leq e^{\theta \rho(t-s)} \frac{1}{M^{\nicefrac{\theta}{d}}(1 - \nicefrac{\theta}{d})}\]
and we can define an MGF-bound for flows following the EBB-model by \(\rho(\theta) = \rho\) and \(\sigma(\theta) = -1/\theta \log(M^{\nicefrac{\theta}{d}}(1 - \nicefrac{\theta}{d}))\). 
\end{example}

The above example is important as it allows us to use results from the tailbounded branch of SNC inside the Calculator. The EBB-model contains important traffic classes such as Markov-modulated On-Off processes (see \cite{Li_eff_bandwidth_2007}).

The next example is influenced by classical queueing theory and is a way to handle the underlying flow being defined in a continuous time setting.

\begin{example}
Assume a Poisson jump process on \(\mathbb{R}\), meaning the interarrival times between any two jumps are independent and exponentially distributed for some intensity parameter \(\mu\). At each jump a packet arrives and the sequence of packets forms the increment process \(a_i\) with \(i \in \mathbb{N}\). The total number of arrivals in a (discrete-timed) interval \((s,t]\) is given by
\[A(s,t) = \sum_{i \in N(s,t)}a_i, \]
where \(N(s,t)\) is the set of jumps which occur in the interval \((s,t]\); now, assume that the increment process \(a_i\) is i.i.d. (for stochastically independent exponential distributions we have the traditional M/M/1-model of queueing theory); then, we can calculate the MGF as
\begin{align*}
    \phi_{A(s,t)}(\theta) & = \sum_{k=0}^\infty \mathbb{E}\left(e^{\theta A(s,t)} \mid N(s,t) = k\right)\mathbb{P}(N(s,t) = k) \\
                          & = e^{\mu(t-s)(\phi_{a_i}(\theta) - 1)};
\end{align*}
hence, the flow is MGF-bounded with \(\rho(\theta) = \mu/\theta (\phi_{a_i}(\theta) - 1)\) and \(\sigma(\theta)=0\).  
\end{example}

In our last example for modeling arrivals we only assume that their distribution are stationary. Instead of having detailed information on their distribution, we model them as the aggregate of sub-flows that (each by their own) have passed through a token bucket shaper.

\begin{example}
Assume a subflow \(A_i\). We say that \(A_i\) has passed through a token bucket shaper, if for all pairs \(s\leq t\) it holds \(A(s,t)\leq \rho_i(t-s) + \sigma_i\). The rate \(\rho_i\) is the shaper's token refreshing rate and \(\sigma_i\) is its bucket size; now, assume that the stochastic processes \(A_i\) are stationary, meaning \(A_i(s,t)\) is equal in distribution to any shift performed to the interval \((s,t)\). For the aggregate \(\sum_i A_i =: A\) the following bound (\cite{Massoulie:tokenbuckets}) holds:
\[\phi_{A(s,t)}(\theta) \leq e^{\theta \sum_i \rho_i(t-s)}\left(1/2 e^{\theta \sum_i \sigma_i} + 1/2 e^{-\theta \sum_i \sigma_i}\right).\]
By defining \(\rho(\theta) = \sum_i \rho_i\) and \(\sigma(\theta) = 1/\theta \log(1/2 e^{\theta \sum_i \sigma_i} + 1/2 e^{-\theta \sum_i \sigma_i})\) we have an MGF-bound in the sense of Definition \ref{def:Arrival-Bound}.
\end{example}

With the above examples we see how to derive MGF-bounds from several models: We covered stochastically independent increments, a conversion from tailbounds, the traditional M/M/1-model, and the result of aggregated shaped traffics. All these bounds are implemented and available in the Calculator.

\subsection{Modeling the Service Process}

The modeling of service elements is usually much easier, as the randomness of service times usually comes from flows interfering with the service element; in fact, the Calculator has currently only one kind of service element implemented, which is the constant rate service.

\begin{example}
We can model a service process \(U\) by a constant rate server, i.e., \(U(s,t) = r(t-s)\) for some rate \(r\). Similarly to the constant rate arrivals the MGF simply is
\[\phi_{U(s,t)}(-\theta) = e^{-r(t-s)}\]
and we achieve an MGF as in Definition \ref{def:Service-Bound} by defining \(\rho(\theta)=-r\) and \(\sigma(\theta)=0\).
\end{example}

We want to briefly point out that, when we deal with a wired system, service elements can usually be modeled as constant rate servers. One should bear in mind, however, that the situation becomes fundamentally different in wireless scenarios. In wireless scenarios, the channel characteristics and properties of wireless nodes  have to be taken into account. More details on the latter can be found in \cite{jiang:servermodel} and \cite{Jiang:IWQoS2010}, where a router's service is parametrized via statistical methods and measurements. This method can likely be applied as a general approach to get a more detailed service description of real-world systems. The modeling of fading channels is addressed in \cite{Fidler:fading-channels}.

In the next section we reason why this simple service model still allows to analyze a wide variety of networks.

\section{Theoretical Results}\label{sec:Theoretical Results}

\subsection{Performance Bounds}

For a system with a single node and a single arrival as in Figure \ref{fig:single-queue}, we have the following performance bounds:
\begin{theorem}\label{thm:Fundamental-Theorem}
Consider the system in Figure \ref{fig:single-queue} and assume that the MGF-bounds 
\begin{align*}
    \phi_{A(s,t)}(\theta) &\leq e^{\theta\rho_A(\theta)(t-s) + \theta\sigma_A(\theta)}\\
    \phi_{U(s,t)}(-\theta) &\leq e^{\theta\rho_U(\theta)(t-s) + \theta\sigma_U(\theta)}
\end{align*}
hold for \(A\) and \(U\) and some \(\theta > 0\). If \(A\) and \(U\) are stochastically independent, then for all \(t>0\) the following bounds hold:
\begin{align*}
    \mathbb{P}(\mathfrak{b}(t)>N) &\leq e^{\theta N}e^{\theta \sigma_A(\theta) + \theta \sigma_U(\theta)} \cdot \frac{1}{1 - e^{\theta (\rho_A(\theta)+\rho_U(\theta))}} \\
    \mathbb{P}(\mathfrak{d}(t)>T) &\leq e^{\theta \rho_U(\theta)T}e^{\theta \sigma_A(\theta) + \theta \sigma_U(\theta)} \cdot \frac{1}{1 - e^{\theta (\rho_A(\theta)+\rho_U(\theta))}},
\end{align*}
if \(\rho_A(\theta)+\rho_U(\theta) < 0\).
\end{theorem}
Proofs for the above theorem can for example be found in \cite{Beck:thesis,fidler-iwqos06}.

This theorem gives us a method to achieve stochastic bounds on the virtual delay and backlog of a single server with a single input. This raises the question on how to achieve performance bounds on more complex networks. The idea here is to reduce a network to the single-flow-single-node case. To illustrate this we give an example of a slightly more complex network.
\begin{example}
Assume the same network as above, but instead of a single flow entering the service element we have two flows \(A_1\) and \(A_2\) as input, each with their own bounding functions \(\rho_i(\theta)\) and \(\sigma_i(\theta)\) (\(i\in\{1,2\}\)). In this scenario we might be interested in the total backlog which can accumulate at the service element. For using the above result, we make an important observation: If \(A_1\) and \(A_2\) are stochastically independent, we can derive from their MGF-bounds a new bound for the aggregated arrivals:
\begin{align*}
    \phi_{A_1(s,t) + A_2(s,t)}(\theta) & =  \mathbb{E}(e^{\theta(A_1(s,t) + A_2(s,t))})=\phi_{A_1(s,t)}(\theta)\phi_{A_2(s,t)}(\theta) \\
                                       & \leq e^{\theta \rho_1(\theta)(t-s) + \theta\sigma_1(\theta)}e^{\theta \rho_2(\theta)(t-s) + \theta\sigma_2(\theta)} \\
                                       & = e^{\theta (\rho_1(\theta) + \rho_2(\theta))(t-s) + \theta(\sigma_1(\theta) + \sigma_2(\theta))}. \\
\end{align*}
By defining \(\rho_A(\theta) = \rho_1(\theta) + \rho_2(\theta)\) and  \(\sigma_A(\theta) = \sigma_1(\theta) + \sigma_2(\theta)\) we can use Theorem \ref{thm:Fundamental-Theorem} again and calculate the aggregated flow's backlog.
\end{example}

It is important to describe exactly what happened in the above example: We have reduced a network consisting of two flows and a service element to a network with only a single flow and a service element. For this we calculated a new MGF-bound consisting of MGF-bounds we have known before. This idea of reducing networks makes SNC a powerful theory.

\subsection{Reduction of Networks}

In this subsection we generalize the above result and show four methods for reducing a network's complexity. The first of these network operations is a repetition of the above example. Proofs and many more details for the following results can for example be found in \cite{Beck:thesis}.
\begin{lemma}\label{lem:Multiplexing}
Assume a service element has two stochastically independent input flows \(A_1\) and \(A_2\) with MGF-bounds \(\rho_i(\theta)\) and \(\sigma_i(\theta)\); then, the aggregate has an MGF-bound with bounding functions \(\rho(\theta) = \rho_1(\theta)+\rho_2(\theta)\) and \(\sigma(\theta) = \sigma_1(\theta) + \sigma_2(\theta)\).
\end{lemma}
The next lemma simplifies a tandem of two service elements into a single service element which describes the end-to-end service.
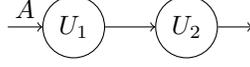
\begin{figure}
    \centering
    \begin{tikzpicture}[scale = 1]   

\path (0,0) node(start) {} 
	++(1,0) node[circle,draw](U_1) {$U_1$} 
	++(1.5,0) node[circle,draw](U_2) {$U_2$} 
	++(1,0) node(end) {}; 
\draw[->] (start) -- node[above] {$A$}(U_1);
\draw[->] (U_1) -- (U_2);
\draw[->] (U_2) -- (end);

\end{tikzpicture}
    \caption{A tandem of two service elements.}
    \label{fig:Tandem-Network}
\end{figure}
\begin{lemma}\label{lem:Convolution}
Assume a tandem network as in Figure \ref{fig:Tandem-Network}, where the processes \(U_1\) and \(U_2\) are stochastically independent and MGF-bounded by the functions \(\rho_i(\theta)\) and \(\sigma_i(\theta)\); then, we can merge the two service elements into a single service element with input \(A\) and output \(C\), representing the end-to-end behavior. It has an MGF-bound with bounding functions \(\rho(\theta) = \max(\rho_1(\theta),\rho_2(\theta))\) and
\[\sigma(\theta) = \sigma_1(\theta) + \sigma_2(\theta) - 1/\theta \log(1 - e^{-\theta |\rho_1(\theta) - \rho_2(\theta)|}).\]
\end{lemma}
The next lemma ties in with a service element's scheduling discipline. Consider again the situation in Figure \ref{fig:single-queue}, but with two input flows \(A_1\) and \(A_2\); now, instead of the system's performance with respect to the aggregated flow we might be interested in the performance for a particular flow. Considering a subflow raises the question of how the flows' arrivals are scheduled inside the service element. From the perspective of SNC the easiest scheduling policy is the strict priority policy (or arbitrary multiplexing): In this policy the flow with lower priority only receives service, if there are no arrivals of the higher priority flow enqueued.
\begin{lemma}\label{lem:Demultiplexing}
Assume the above described scenario and that \(A_1\) and \(U\) are stochastically independent with bounding functions \(\rho_A(\theta),\sigma_A(\theta)\) and \(\rho_U(\theta),\sigma_U(\theta)\), respectively. This system can be reduced to a single-flow-single-node system for flow \(A_2\) and a service element \(U_l\) with MGF-bound
\[\phi_{U_l(s,t)}(-\theta) \leq e^{\theta(\rho_A(\theta)+\rho_U(\theta))(t-s) + \theta(\sigma_A(\theta)+\sigma_U(\theta))}.\]
\end{lemma}
More elaborate scheduling policies have been analyzed in SNC. At this stage, however, the Calculator has only implemented the above method for calculating leftover service. Note that this is a worst case view with respect to the scheduling algorithm. By this we mean that any other scheduling, like FIFO or WFQ, gives more service to \(A_2\) than arbitrary multiplexing does; therefore, the result of the above lemma can always be used as a lower bound for the service \(A_2\) receives.

The next result is needed to produce results for intermediate nodes or flows. It gives an MGF-bound for a service element's output.

\begin{lemma}\label{lem:Deconvolution}
Assume the scenario as in Figure \ref{fig:single-queue} again. If \(A\) and \(U\) are stochastically independent and MGF-bounded by bounding functions \(\rho_A(\theta),\sigma_A(\theta)\) and \(\rho_U(\theta),\sigma_U(\theta)\), respectively, we have for the output flow \(B\):
\[\phi_{B(s,t)}(\theta) \leq e^{\theta\rho_A(\theta)(t-s) + \theta(\sigma_A(\theta)+\sigma_U(\theta))}\cdot \frac{1}{1 - e^{\theta(\rho_A(\theta)+\rho_U(\theta))}},\]
if \(\rho_A(\theta)+\rho_U(\theta) < 0\).
By this \(B\) is MGF-bounded with \(\rho_B(\theta) = \rho_A(\theta)\) and \[\sigma_B(\theta) =\sigma_A(\theta) + \sigma_U(\theta) - 1/\theta \log(1 - e^{\theta(\rho_A(\theta) + \rho_U(\theta))}).\]
\end{lemma}

All of the above results required some independence assumption between the analyzed objects. For the analysis of stochastically dependent objects, we use H\"{o}lder's inequality:

\begin{lemma}
Let \(X\) and \(Y\) be two stochastic processes. It holds
\[\mathbb{E}(XY) \leq (\mathbb{E}(X^p))^{\nicefrac{1}{p}}(\mathbb{E}(Y^q))^{\nicefrac{1}{q}}\]
for all pairs \(p,q\) such that \(\tfrac{1}{p} + \tfrac{1}{q} = 1\). In particular we have
\[\phi_{XY}(\theta) \leq (\phi_{X}(p\theta))^{\tfrac{1}{p}}(\phi_{Y}(q\theta))^{\tfrac{1}{q}}.\]
\end{lemma}

When we apply this inequality to the above results we get a modified set of network operations. For more details, we refer again to \cite{Beck:thesis}.

\begin{lemma}
In the case of stochastic dependence the bounding functions in Lemma \ref{lem:Multiplexing} change to \(\rho(\theta) = \rho_1(p\theta) + \rho_2(q\theta)\) and \(\sigma(\theta) = \sigma_1(p\theta) + \sigma_2(q\theta)\).
\end{lemma}
\begin{lemma}\label{lem:Dependent-Convolution}
In the case of stochastic dependence the bounding functions in Lemma \ref{lem:Convolution} change to \(\rho(\theta) = \max(\rho_1(p\theta) + \rho_2(q\theta))\) and 
\[\sigma(\theta) = \sigma_1(p\theta) + \sigma_2(q\theta) - 1/\theta \log(1 - e^{-\theta |\rho_1(p\theta) - \rho_2(q\theta)|}).\]
\end{lemma}
\begin{lemma}
In the case of stochastic dependence the bounding functions in Lemma \ref{lem:Demultiplexing} change to \(\rho(\theta) = \rho_A(p\theta) + \rho_U(q\theta)\) and \(\sigma(\theta) = \sigma_A(p\theta) + \sigma_U(q\theta)\).
\end{lemma}
\begin{lemma}
In the case of stochastic dependence the bounding functions in Lemma \ref{lem:Deconvolution} change to \(\rho_B(\theta) = \rho_A(p\theta) \) and 
\[\sigma_B(\theta) =\sigma_A(p\theta) + \sigma_U(q\theta) - 1/\theta \log(1 - e^{\theta(\rho_A(p\theta) + \rho_U(q\theta))}).\]
\end{lemma}

Now, we show how these network operations work together to reduce a complex network to the single-node-single-flow case. These examples are taken directly from Chapter 1 of \cite{Beck:thesis} lifted to MGF-bounded calculus.

\begin{figure}
    \centering
    \begin{tikzpicture}[scale = 1]   

\path (-1,0) node(label) {$\mathcal{G}:$}
	  (0,0.25) node(Origin_1) {}
	++(1,0) node(dummy_U_top)[text = white]{$\cdots$}
	++(0,-0.25) node(U)[circle, draw]{$U$}
	++(1,0.25) node(dummy_V_top)[text = white]{$\cdots$}
	++(0,-0.25) node(V)[circle, draw]{$V$}
	++(1,0.25) node(Destination_1) {}
	  (0,-0.25) node(Origin_2) {}
	++(1,0) node(dummy_U_bottom)[text = white]{$\cdots$}
	++(1,0) node(dummy_V_bottom)[text = white]{$\cdots$}
	++(1,0) node(Destination_2) {};

\draw[->] (Origin_1) -- node[above] {${A}_1$}(dummy_U_top); 
\draw[->] (dummy_U_top) -- (dummy_V_top);
\draw[->] (dummy_V_top) -- (Destination_1);

\draw[->] (Origin_2) -- node[below] {${A}_2$}(dummy_U_bottom); 
\draw[->] (dummy_U_bottom) -- (dummy_V_bottom);
\draw[->] (dummy_V_bottom) -- (Destination_2);

\end{tikzpicture}
    \caption{A two nodes and two flows example. Our goal is to reduce this network to the single node single flow case.}
    \label{fig:2-Nodes-2-Flows}
\end{figure}
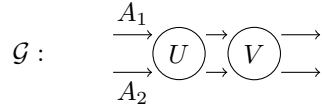

We consider the network of Figure \ref{fig:2-Nodes-2-Flows} and assume that the following MGF-bounds on the involved elements hold:
\begin{align*}
    \phi_{A_1(s,t)}(\theta) & \leq e^{\theta\rho_{A_1}(\theta)(t-s) + \theta\sigma_{A_1}(\theta)} \\
    \phi_{A_2(s,t)}(\theta) & \leq e^{\theta\rho_{A_2}(\theta)(t-s) + \theta\sigma_{A_2}(\theta)} \\
    \phi_{U(s,t)}(\theta) & \leq e^{\theta\rho_{U}(\theta)(t-s) + \theta\sigma_{U}(\theta)} \\
    \phi_{V(s,t)}(\theta) & \leq e^{\theta\rho_{V}(\theta)(t-s) + \theta\sigma_{V}(\theta)}.
\end{align*}

We present three examples for reducing the network using the operations defined in the lemmata above. 

\begin{figure}
    \centering
    \begin{tikzpicture}[scale = 1]   

\path (-1,0) node(label) {$\mathcal{G}_1:$}
	++(1.5,0) node(Origin) {${A}_1 \oplus {A}_2$}
	++(2.5,0) node(Service)[ellipse, draw]{$U\otimes V$};

\draw[->] (Origin) -- (Service); 
\end{tikzpicture}
\break
\medskip
(a) Convolution after multiplexing \break
\medskip
\begin{tikzpicture}[scale = 1]   

\path (-1,0) node(label) {$\mathcal{G}_1^\prime:$}
	++(2,0) node(Origin) {$({A}_1 \oplus {A}_2)\oslash U$}
	++(2,0) node(Service)[circle, draw]{$V$};

\draw[->] (Origin) -- (Service); 

\end{tikzpicture} \break
(b) Deconvolution after multiplexing.\break
    \caption{Resulting graphs for aggregating first.}
    \label{fig:Reduction-Example-Aggregate-First}
\end{figure}
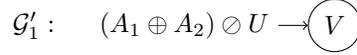

\begin{example}
Consider the graph \(\mathcal{G}\) given in Figure \ref{fig:2-Nodes-2-Flows}. After merging both arrivals the graph can be simplified in two ways: Either apply Lemma \ref{lem:Convolution} to the two service elements (resulting in graph \(\mathcal{G}_{1}\) in Figure \ref{fig:Reduction-Example-Aggregate-First}) or calculate an output bound for the first node's departures (resulting in graph \(\mathcal{G}_{1}^{\prime}\) in Figure \ref{fig:Reduction-Example-Aggregate-First}). The graphs \(\mathcal{G}_{1}\) and \(\mathcal{G}_{1}^{\prime}\) describe the system for both arrivals aggregated and as such, can also be used to calculate performance bounds for only one of the flows. The difference between these two methods is that the graph \(\mathcal{G}_{1}\) describes the system's end-to-end behavior, whereas \(\mathcal{G}_{1}^{\prime}\) describes the behavior at the service element \(V\). The MGF-bounds of the quantities appearing in Figure \ref{fig:Reduction-Example-Aggregate-First} can be calculated using Lemmas \ref{lem:Multiplexing}-\ref{lem:Deconvolution}. To show how these work together we derive here the bounding functions for the MGF-bound on \(A^\prime := (A_1\oplus A_2)\oslash U\): First we combine the MGF-bounds of \(A_1\) and \(A_2\) into the MGF-bound
\[\phi_{A_1(s,t)+A_2(s,t)}(\theta) \leq e^{\theta(\rho_{A_1}(\theta)+\rho_{A_2}(\theta))(t-s) + \theta(\sigma_{A_1}(\theta)+\sigma_{A_2}(\theta))}.\]
Next, we apply Lemma \ref{lem:Deconvolution} to the aggregate and the service process \(U\), resulting in 
\[\phi_{A^\prime(s,t)}(\theta) \leq e^{\theta(\rho_{A_1}(\theta)+\rho_{A_2}(\theta))(t-s) + \theta(\sigma_{A_1}(\theta)+\sigma_{A_2}(\theta))} \frac{1}{1 - e^{\theta(\rho_{A_1}(\theta)+\rho_{A_2}(\theta)+\rho_U(\theta))}},\]
if \(\rho_{A_1}(\theta)+\rho_{A_2}(\theta)+\rho_U(\theta)<0\).
\end{example}

\begin{figure}
    \centering
    \begin{tikzpicture}[scale = 1]   

\path (-1,0) node(label) {$\mathcal{G}_2:$}
	++(1,0) node(Origin) {${A}_1$}
	++(4.5,0) node(Service)[ellipse, draw]{$[U\ominus{A}_2]^+\otimes[V\ominus({A}_2\oslash U)]^+$};

\draw[->] (Origin) -- (Service); 

\end{tikzpicture}\break
\medskip
(a) Convolution after subtraction.\break
\medskip
\begin{tikzpicture}[scale = 1]   

\path (-1,0) node(label) {$\mathcal{G}_2^\prime:$}
	++(2,0) node(Origin) {${A}_1 \oslash [U\ominus {A}_2]^+$}
	++(4,0) node(Service)[ellipse, draw]{$[V\ominus({A}_2\oslash U)]^+$};

\draw[->] (Origin) -- (Service); 

\end{tikzpicture}\break
\medskip
    \caption{Resulting graphs for subtracting first.}
    \label{fig:Reduction-Example-Subtract-First}
\end{figure}
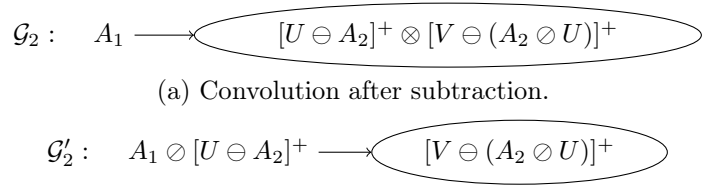

\begin{example}\label{ex:Subtract-First}
Another method to reduce \(\mathcal{G}\) is to subtract one of the flows -- say \(A_{2}\) -- first. Afterwards either Lemma \ref{lem:Convolution} or Lemma \ref{lem:Deconvolution} can be applied, leading to the graphs \(\mathcal{G}_{2}\) and \(\mathcal{G}_{2}^{\prime}\) in Figure \ref{fig:Reduction-Example-Subtract-First}. The graph \(\mathcal{G}_{2}\) describes an end-to-end behavior, whereas \(\mathcal{G}_{2}^{\prime}\) is the local analysis at the second node. In contrast to the previous example, the flows are considered separately throughout the whole analysis. This approach proves to be better in general topologies in which flows interfere only locally. Note that by following this approach there occurs a stochastic dependency in graph \(\mathcal{G}_2\) when we use \ref{lem:Convolution}: The process \(A_2\) appears in both service descriptions. As a consequence we need to use its variation formulated in Lemma \ref{lem:Dependent-Convolution}, which introduces a set of H\"{o}lder parameters; similarly, in graph \(\mathcal{G}^\prime_2\) we have to employ a variation of Theorem \ref{thm:Fundamental-Theorem} when we want to calculate performance bounds (the process \(A_2\) appears in the arrivals and in the service description).
\end{example}

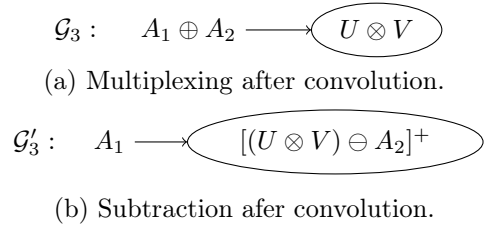
\begin{figure}
    \centering
    \begin{tikzpicture}[scale = 1]   

\path (-1,0) node(label) {$\mathcal{G}_3:$}
	++(1.5,0) node(Origin) {${A}_1 \oplus {A}_2$}
	++(2.5,0) node(Service)[ellipse, draw]{$U\otimes V$};

\draw[->] (Origin) -- (Service); 

\end{tikzpicture}\break
\medskip
(a) Multiplexing after convolution.\break
\medskip
\begin{tikzpicture}[scale = 1]   

\path (-1,0) node(label) {$\mathcal{G}_3^\prime:$}
	++(1,0) node(Origin) {${A}_1$}
	++(3,0) node(Service)[ellipse, draw]{$[(U\otimes V)\ominus {A}_2]^+$};

\draw[->] (Origin) -- (Service); 

\end{tikzpicture}\break
\medskip
(b) Subtraction afer convolution.
    \caption{Resulting graphs for conlvolving first.}
    \label{fig:Reduction-Example-Convolve-First}
\end{figure}

\begin{example}
Instead of merging one of the edges first, one can also use Lemma \ref{lem:Convolution} to merge the two service elements first. The resulting node is labeled by \(U\otimes V\). The graph \(\mathcal{G}_{3}\) in Figure \ref{fig:Reduction-Example-Convolve-First}(a) equals \(\mathcal{G}_{1}\); indeed, just the order of aggregation and convolution was switched. Subtracting a crossflow from the convoluted service element, instead, would lead to Figure \ref{fig:Reduction-Example-Convolve-First}(b). This last graph \(\mathcal{G}^\prime_3\) is generally assumed to yield the best end-to-end bounds for flow the flow \(A_1\); however, this strategy of convoluting before calculating leftover services cannot be applied in general feedforward networks.
\end{example}

We see there are several ways of reducing even this simple example of a network. The results differ in quality and also in what exactly we want to analyze (the performance with respect to a single flow vs. the aggregate and the end-to-end performance vs. the performance at the network's second node). Note also that the choice of network operations applied may or may not result in H\"{o}lder parameters appearing in the resulting performance bounds; therefore any automatic process that performs these actions must keep track whether stochastic dependencies occur and where exactly H\"{o}lder parameters must be introduced.

\subsection{End-to-end Results}
Now, we discuss SNC's capabilities to perform an end-to-end analysis of a queueing system. Often one is interested in the end-to-end delay of a tandem of servers as in Figure \ref{fig:Tandem-Network}, but with \(n\) service elements instead of two. A typical scenario would be the end-to-end delay between a client and a server with several routers or switches in between. 

Given such a network we could theoretically calculate an end-to-end delay bound in two ways: (1) We could start by reducing the network to the first service element and calculate a delay bound for this element in isolation; next, we reduce the original network to the second service element and calculate another local delay bound and so on. All these single-node delay bounds can be combined into an end-to-end delay bound by ``adding them up''. While this approach works in theory, we know the resulting bounds to be very loose in general. (2) The other approach is to use Lemma \ref{lem:Convolution} to get an end-to-end description for the system and use it to derive delay bound directly. From the theory of network calculus we know that this approach is beneficial.

Still, in this second course of action there exists a problem: Inspecting Lemma \ref{lem:Convolution} reveals that with each application of it a term of the form \(\frac{1}{1 - e^{\theta |\rho_i(\theta) - \rho_{i+1}(\theta)|}}\) enters the equations. These terms worsen the delay bounds, especially when the quantitites \(\rho_i(\theta)\) and \(\rho_{i+1}(\theta)\) are similarly sized (in fact, if they should be equal the lemma cannot deliver this result at all). The next theorem shows a method for avoiding these terms completely (see Theorem 3.1 in \cite{Beck:thesis} and also \cite{fidler-iwqos06}). This can be seen as an end-to-end convolution, whereas the successively applying Lemma \ref{lem:Convolution} would compare to a node-by-node convolution.

\begin{theorem}\label{thm:End-to-End}
Fix some \(\theta>0\) and consider a sequence of two service elements as in Lemma \ref{lem:Convolution}; further, let \(A\) be MGF-bounded with functions \(\rho_A\) and \(\sigma_A\). Let \(A\), \(U\), and \(V\) be stochastically independent. Under the stability condition \break\(\rho_{A}(\theta)<-\rho_{U}(\theta)\wedge-\rho_{V}(\theta)\), the end-to-end performance bounds 
\begin{align*}
    \mathbb{P}(\mathfrak{d}(t)>T)\leq e^{-\theta \rho_A(\theta)T} \frac{e^{\theta(\sigma_{A}(\theta)+\sigma_{U}(\theta)+\sigma_{V}(\theta))}} {(1-e^{\theta(\rho_{U}(\theta)+\rho_{A}(\theta))})(1-e^{\theta(\rho_{V}(\theta)+\rho_{A}(\theta))})}
\end{align*}
holds for all \(t\) and \(T\).
\end{theorem}
Above theorem easily generalizes to \(N\) hops. Denoting the bounding functions of the \(i\)-th server by \(\rho_{i}\) and \(\sigma_i\) we have
\[\mathbb{P}(\mathfrak{d}(t)>T)\leq e^{-\theta \rho_A(\theta)T} \frac{e^{\theta \sigma_A(\theta) + \sum_i \theta \sigma_i(\theta)}} {\prod_i 1 - e^{\theta(\rho_i(\theta) +\rho_A(\theta))}}\]
under the stability condition \(\rho_A(\theta) < \bigwedge_i - \rho_i(\theta)\).

For stochastically dependent services or arrivals, the introduction of H\"{o}lder parameters is needed similarly to the previous subsection.

Note that by using lemmata \ref{lem:Multiplexing} - \ref{lem:Deconvolution} (or their respective variants for stochastically dependent cases) we can reduce any feedforward network to a tandem of \(N\) service elements for any flow of interest with \(N\) hops. In doing so, however, the exact sequence of performed network operations will determine the number of H\"{o}lder parameters. The optimal way of reducing the network to the tandem is not known and subject to current research.

\section{Code Structure: An Overview on the Calculator}\label{sec:Code Structure}

Now, we give an overview on the Calculator and its code structure. The work-flow with the program is the following.
\begin{enumerate}
    \item The network must be modeled and given to the Calculator by the user. This requires deriving MGF-bounds for all input-flows and service-elements. If pre-existing stochastic dependencies are known, they must be given to the program. Otheriwse the program will assume stochastic independence. This does only include the initial stochastic processes given to the program, for any intermediate results the tool will keep track of stochastic dependencies by itself. For example the stochastic dependencies occurring in Example \ref{ex:Subtract-First} will be recognized by the tool.
    Networks can either be entered through the GUI or by loading a file holding the description.
    \item After giving the network to the tool it can perform an \emph{analysis} of it for any given flow and node of interest (or flow and path of interest). This translates into using Lemmata \ref{lem:Multiplexing} - \ref{lem:Deconvolution} (and their variants) until the network has been reduced to one on which Theorem \ref{thm:Fundamental-Theorem} can be applied. This step is performed entirely on a symbolic level, meaning: The Calculator works on the level of functions and composes these as defined by Lemmata \ref{lem:Multiplexing} - \ref{lem:Deconvolution}. As a last action of this analysis step the tool applies Theorem \ref{thm:Fundamental-Theorem} (again on the level of functions) to compute the function that describes the delay- or backlog bound.
    \item In the \emph{optimization}-step the tool takes the function from the analysis-step and optimizes it by all the parameters it includes. The parameters to be optimized include at least \(\theta\) and might include any number of additional H\"{o}lder parameters; consequently, it is important that any optimization-method implemented to the tool is flexible to the actual number of optimization parameters occurring. This step is usually the one that takes the most computational time.
\end{enumerate}

\begin{figure}
    \centering
    \begin{tikzpicture}[scale = 1]   

\path (0,0) node[ellipse, draw](User) {User}
	++(0,-1) node[rectangle, draw](GUI) {GUI}
	++(0,-1) node[rectangle, draw](Calculator) {Calculator}
	++(-4,-1) node(Network) {Network}
	++(0,-1) node[rectangle, draw](Flow) {Flows}
	++(0,-1) node[rectangle, draw](Hoelder) {Hoelder}
	++(0,-1) node[rectangle, draw](Vertex) {Vertices}

	++(4,3) node[rectangle, draw](Analysis) {Analysis}
	++(4,0) node[rectangle, draw](Optimizer) {Optimizer};

\draw[<->] (User) -- (GUI); 
\draw[<->] (GUI) -- (Calculator);
\draw[<->] (Calculator) -- (-3,-2.85);
\draw (-5,-2.75) rectangle (-3,-6.3);
\draw[<->] (Calculator) -- (Analysis);
\draw[<->] (Calculator) -- (Optimizer);

\end{tikzpicture}
    \caption{Relation of the Calculator's components.}
    \label{fig:Calculator-Modules}
\end{figure}
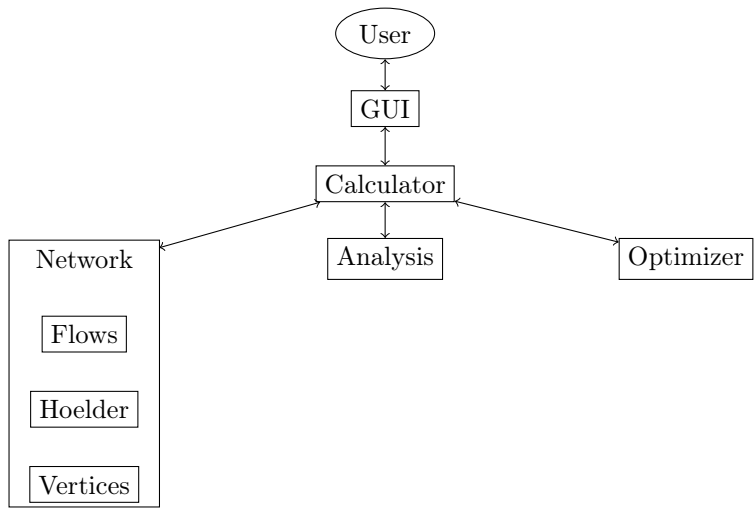

The tool reflects the above roadmap by consisting of different exchangeable parts. Figure \ref{fig:Calculator-Modules} presents these modules.
\begin{itemize}
    \item The GUI is the interface between the program's core and the user. We have implemented a simple GUI for the tool, which allows to construct and manipulate the network by hand. It also gives access to the program's analysis-part and optimization-part. Note that the GUI is not necessary to use the calculator, instead, the provided packages can be mostly used like a library.
    \item The Calculator is the core of the program. It is the interface between the other models and relays commands and information as needed.
    \item The Network stores all the needed topology. This includes the flows and nodes with their parameters, but also MGF-bounds on service processes and flows and H\"{o}lder parameters that are created during the analysis-step.
    \item The Analysis is responsible to performing the algebraic part. It is coded entirely on a symbolic level. 
    \item The Optimizer has the task to ``fill'' the functions given by the Analysis with numerical values. Following an optimization strategy (or heuristic) it will find a near optimal set of parameters and calculate the corresponding performance bound.
\end{itemize}

The calculator's main class is the \code{SNC} class, which is a singleton, bridging the communication between GUI and backend. Alternatively, the \code{SNC} class can be used directly. It provides a command-pattern-style interface; accordingly, the underlying network is altered through sending commands to the \code{SNC} class; moreover, these commands are stored in an \code{undoRedoStack}, which is be used to track, undo and redo changes.
Note that loosening connections and facilitating the use of the backend without a GUI is still work in progress.

When exceptions occur in the backend, the control flow will try to repair things as good as possible. If this is not possible, a generic runtime exception is thrown. These exceptions are specified in \code{misc} at the moment. The design choice for generic exceptions extending the java build-in \code{RuntimeException} stems from the fact that we wanted to avoid methods with a variety of checked exceptions. As a result, the code is less cluttered and more readable. We are aware that this topic is under debate, especially in the java community and that this practice needs thorough documentation.

Currently the code is organized into the following packages all packages start with \code{unikl.disco.}, which we omit in this list:
\begin{itemize}
    \item \code{calculator} This package contains only the main class, called \code{SNC}. It is the core of the program.
    \item \code{calculator.commands} This package contains the various commands by which the network is manipulated (adding a flow, removing a vertex, etc.)
    \item \code{calculator.gui} This package includes all classes needed to generate, display, and interact with the tool's GUI. Except for the \code{FlowEditor} class, the GUI is modular, easy to change and extend since actions are separated from markup.
    \item \code{calculator.network} This package includes all classes related to topological information like \code{Flow} and \code{Vertex}. It further contains the classes needed to perform an analysis.
    \item \code{calculator.optimization} This class contains all classes needed to evaluate and optimize performance bounds. Note that the result of an analysis is given as an object of type \code{Flow} and belongs to the \code{calculator.network} package. The bound's information must be ``translated'' into backlog- and delay bounds first, before numerical values can be provided. This is why we find in the \code{optimization}-package classes like \code{BacklogBound} or \code{BoundType}.
    \item \code{calculator.symbolic\_math} This is a collection of algebraic manipulations. We find for example an \code{AdditiveComposition} here, which just combines two symbolic functions into a new one. When evaluated with a set of parameters this returns the sum of the two atom-functions evaluated with the same set of parameters; furthermore, we find in this package symbolic representations of MGF-bounds for flows (\code{Arrival}) and service elements (\code{Service}).
    \item \code{calculator.symbolic\_math.functions} Some arrival classes or specific manipulations (such as Lemma \ref{lem:Deconvolution}) require the repeated usage of very specific algebraic manipulations. This package collects these operations.
    \item \code{misc} A package containing miscellaneous classes not fitting well anywhere else and generic runtime exceptions.
\end{itemize}

\section{Input \& Output}
\label{sec:input_output}

There are several different methods to input and output data when using the calculator:
\begin{itemize}
    \item Using the functions provided by the GUI
    \item Writing/reading networks from/to text files
    \item Using custom defined functions when the calculator is used as a library
\end{itemize}

The GUI provides methods for adding, removing, subtracting and convoluting flows and vertices. This is useful for quick tests and experimentation but cumbersome for larger networks. To that end, we implemented a simple text-file based interface for saving and loading networks. Note that at the moment only networks without dependencies can be loaded/saved! Extending these methods to the general case is subject to future work.

The file format is specified as follows:
\begin{itemize}
    \item A line starting with ``\#'' is a comment and ignored
    \item First, the interfaces (vertices) are specified, each in its own line
    \item After the last interface definition, an ``EOI'' ends the interface block
    \item Then, the flows are specified, again, each in its own line, until ``EOF'' ends the document
\end{itemize}

An interface line has the following syntax (without the linebreak):
\begin{verbatim}
I <vertexname>, <scheduling policy>, <type of service>, 
<parameters> ...
\end{verbatim}

At the moment only FIFO scheduling and constant rate (``CR'') service are supported. A flow line has the following syntax (again, without the linebreak):
\begin{verbatim}
F <flowname>, <number of vertices on route>, 
<name of first hop>:<priority at this hop>, ..., 
<type of arrival at first hop>, <parameters> ...
\end{verbatim}

The priority is a natural number with 0 being the highest. At the moment, the following arrival types are possible (with parameters):
\begin{itemize}
    \item CONSTANT -- service rate ($\geq 0$!)
    \item EXPONENTIAL -- mean 
    \item EBB (exponentially bounded burstiness) -- rate, decay, prefactor
    \item STATIONARYTB -- rate, bucket, [maxTheta]
\end{itemize}

where \code{maxTheta} is optional. In the following we see a sample network with three vertices and one exponentially distributed flow.

\begin{verbatim}
# Configuration of a simple network
# Interface configuration. Unit: Mbps
I v1, FIFO, CR, 1
I v2, FIFO, CR, 3
I v3, FIFO, CR, 4

EOI
# Traffic configuration. Unit Mbps or Mb
# One flow with the route v1->v2->v3
F F1, 3, v1:1, v2:1, v3:2, EXPONENTIAL, 2
EOF
\end{verbatim}

\section{Code Representation of SNC Results and Concepts}\label{sec:Code Representation}

Now we elaborate on some of the most important classes of the Calculator and how they represent core-concepts of SNC.
\subsection{The \code{Network}-class}
This class is responsible for storing a network's topology and manipulating its elements. Its key members are three \code{Map}s:
\begin{itemize}
    \item \code{flows} is of type \code{Map<Integer,Flow>} and is a collection of flows, each with a unique ID. Each flow represents one flow's entire path through the network. See also the subsection about \code{flows} below.
    \item \code{vertices} is of type \code{Map<Integer,Vertex>} and is a collection of vertices, each with a unique ID. Each vertex represents one service element of the network. It does not matter how many flows this service element has to process, it will always be modeled by a single \code{Vertex}. See also the below subsection about \code{vertices}.
    \item \code{hoelders} is of type \code{Map<Integer,Hoelder>}. Each newly introduced H\"{o}lder parameter (actually the pair of parameters \(p\) and \(q\) are defined by \(1/p +1/q = 1\) and can be represented by a single variable) is collected in this object and receives a unique ID. This data-structure is needed to keep track of and distinguish the introduced parameters.
\end{itemize}
These \code{Map}s are created and manipulated by various methods of the \code{Network}-class. Some of these methods are straightforward such as \code{addVertex}, \code{addFlow}, and \code{setServiceAt}. Others methods are more involved and directly reflect core concepts of SNC: The method \code{computeLeftover} for example manipulates the network like follows: For a specific node it identifies the flow that has priority at this service element. It then calculates the leftover service description after serving this flow (Lemma \ref{lem:Demultiplexing}) and gives this description to this \code{Vertex}; furthermore, the method gives as output an object of type \code{Arrival}, which encodes the MGF bound on this node's output for the just served flow (Lemma \ref{lem:Deconvolution}).

\subsection{The \code{Flow}-class and the \code{Arrival}-class}

These two classes are closely related: The \code{Flow} class can be thought of as the topological information of a flow through the network. It contains a \code{List} of integer-IDs that describes the flow's path through the network and a \code{List} of corresponding priorities. It further has a \code{List} of \code{Arrival}-objects. These objects describe the flow's MGF-bounds at a given node. Usually a flow added to the network only has a single \code{Arrival}-object in this list, which is the MGF-bound at that flow's the ingress node. Every \code{Flow}-object keeps track of for how many hops arrival-bounds are known in the integer variable \code{established\_arrivals}.

An \code{Arrival}-object most important members are the two \code{SymbolicFunction}s \code{rho} and \code{sigma}. These directly represent the bounding-functions \(\rho\) and \(\sigma\) of an MGF-bound (see Definition \ref{def:Arrival-Bound}); further, important members are two\break\code{Set<Integer>}s, which keep track of the flows' and services' IDs this arrival is stochastically dependent to, respectively.

\subsection{The \code{Vertex}-class and the \code{Service}-class}

Similarly to \code{Flow} and \code{Arrival} these two classes are closely connected to each other. Each \code{Vertex}-object has a member of type \code{Service}, which describes its service via an MGF-bound (Definition \ref{def:Service-Bound}); furthermore, a \code{Vertex}-object has members \code{priorities} (of type\break \code{Map<Integer,Integer>}) and \code{incoming} (of type \code{Map<Integer, Arrival>}) to identify which flow would receive the full service and what set of flows are incoming to this node.

An \code{Service}-object is the equivalent of an \code{Arrival}-object on the service side. It also contains two \code{SymbolicFunction}s called \code{rho} and \code{sigma} and two \code{Set<Integer>} to keep track of stochastic dependencies.

\subsection{The \code{SymbolicFunction}-interface}

This interface lies at the core of the symbolic computations made to analyze a network. Each MGF-bound is represented by two functions \(\rho\) and \(\sigma\), which find their representation as \code{SymbolicFunction} in the code. This interface's most important method is the \code{getValue}-method. It takes a \code{double} (the \(\theta\)) and a \code{Map<Integer,Hoelder>} (the -- possibly empty -- set of H\"{o}lder parameters) as input and evaluates the function at this point. A simple example is the \code{ConstantFunction}-class, which implements this interface. When the method \code{getValue} is called, an object of this kind just returns a constant value, given that the \code{Map<Integer,Hoelder>} was empty. Mathematically written such an object just represents \(f(\theta) = c\), which can for example be found in the MGF-bound of constant rate arrivals or service elements.

The modeling power here lies in the composition of \code{SymbolicFunctions}; for example, when we want to merge two constant rate arrivals, their MGF-bound would contain \(\rho_{agg}(\theta) = r_1 + r_2\) with \(\rho_1(\theta) = r_1\) and \(\rho_2(\theta)= r_2\) being the subflows' rates, respectively. The class \code{AdditiveComposition} implements the \code{SymbolicFunction}-interface itself and has two members of type \code{SymbolicFunction}. These are called atom-functions; in this scenario the atom-functions would be two\break \code{ConstantFunction}s with rates \(r_1\) and \(r_2\). When the \code{getValue}-method of\break \code{AdditiveComposition} is called it will relay the given parameters to its atom-functions and get their values (\(r_1\) and \(r_2\)) and return their sum to the caller; indeed, \code{AdditiveComposition} is just a representation of the plus-sign in\break \(r_1 + r_2 = \rho_1(\theta) + \rho_2(\theta)\).

\subsection{The \code{AbstractAnalysis}-class}

The abstract class \code{AbstractAnalysis} defines the methods and members an analysis of a network needs. It serves as a starting point for concrete analysis classes. Its members include a network's topological information together with the indices of the flow of interest and the service of interest, so the analysis knows what performance the caller is interested in. The important method here is the \code{analyze}-method of the \code{Analyzer} interface, which every Analysis has to implement. This method (to be defined by any realization of this abstract class) gives as output an object of type \code{Arrival}, which represents a performance bound; in fact, remembering the bounds from Theorem \ref{thm:Fundamental-Theorem}
\begin{align*}
    \mathbb{P}(\mathfrak{b}(t)>N) &\leq e^{\theta N}e^{\theta \sigma_A(\theta) + \theta \sigma_U(\theta)} \cdot \frac{1}{1 - e^{\theta (\rho_A(\theta)+\rho_U(\theta))}} \\
    \mathbb{P}(\mathfrak{d}(t)>T) &\leq e^{\theta \rho_U(\theta)T}e^{\theta \sigma_A(\theta) + \theta \sigma_U(\theta)} \cdot \frac{1}{1 - e^{\theta (\rho_A(\theta)+\rho_U(\theta))}}.
\end{align*}
we see that we can split these bounds into a part that depends on the bound's value (\(N\) or \(T\), respectively) and a factor that does not depend on the bound's value. So, we can also write:
\[\mathbb{P}(\mathfrak{b}(t)>N) \leq e^{\theta \rho_\mathfrak{b}(\theta) N + \theta \sigma_\mathfrak{b}(\theta)}\]
with \(\rho_\mathfrak{b}(\theta) := 1\) and \(\sigma_\mathfrak{b} := \sigma_A(\theta) + \theta_U(\theta) - 1/\theta \log(1 - e^{\theta (\rho_A(\theta) + \rho_U(\theta))})\). And:
\[\mathbb{P}(\mathfrak{d}(t)>T) \leq e^{\theta \rho_\mathfrak{d}(\theta) T + \theta \sigma_\mathfrak{d}(\theta)}\]
with \(\rho_\mathfrak{d}(\theta) = \theta \rho_U(\theta)\) and \(\sigma_\mathfrak{d} = \sigma_\mathfrak{b}\). This representation has the advantage that we can use the already implemented operations for MGF-bounds for our performance bounds; for this reason, the output of the \code{analyze}-method is an object of type \code{Arrival}, which is how the code represents an MGF-bound.

\subsection{The \code{AbstractOptimizer}-class}

Similar to the \code{AbstractAnalysis} class, this abstract class serves as a starting point for implementing optimizers. It implements the \code{minimize}-method of the \code{Optimizer} interface, which every optimizer has to implement. This method takes as input the granularity for which the continuous space of optimization parameters should be discretized too and returns the minimal value found by the optimization algorithm. Its most important member is \code{bound}, which is basically the MGF-bound presented in the previous subsection. Together with the class \code{BoundType} and the interface \code{Optimizable} the function to be optimized is defined. This can either be a backlog- or delay bound as defined in the previous subsection or it can be their inverse functions, i.e., the smallest bound \(N\) or \(T\) that can be found for a given violation probability \(\varepsilon\). for this the bounds from Theorem \ref{thm:Fundamental-Theorem} must be solved for \(N\) and \(T\).

\section{APIs and Extending the Calculator}\label{sec:APIs and Extensions}

The interfaces and abstract classes provide a good starting point when extending the calculator. The backend makes heavy use of the factory pattern. As long as new classes implement the necessary interfaces, extending the behavior is easy. The only exception being the \code{FlowEditor} of the GUI, which we are planning to rewrite as soon as possible.

In this section we describe in more detail how a user can implement his own models into it. For this we cover four cases; we describe how users can implement their own
\begin{itemize}
    \item arrival model to the Calculator and its GUI, given some known MGF-bounds.
    \item service model to the Calculator and its GUI, given some known MGF-bounds.
    \item method of analysis to the Calculator and its GUI.
    \item method of parameter optimization to the Calculator and its GUI. 
\end{itemize}
These descriptions are subject to changes of the code and we strongly recommend to pay attention to the code's documentation before implementing any of the above.

\subsection{Adding Arrival Models}
To add a new arrival model to the calculator we need to be able to write the arrivals in an MGF-bounded form as in Definition \ref{def:Arrival-Bound}. We consider for this documentation an arrival that has an exponential amount of data arriving in each time step with rate parameter \(\lambda\) as example (see\break Example \ref{ex:Exponential-Increments}): \[\mathbb{E}(e^{\theta(A(t)-A(s))})\leq\left(\frac{\lambda}{\lambda-\theta}\right)^{t-s}\qquad\text{for all }\theta<\lambda.\]
In this case \(\rho(\theta)=\tfrac{1}{\theta}\log(\tfrac{\lambda}{\lambda-\theta})\) and \(\sigma(\theta)=0\).

When we have appropriate \(\sigma\) and \(\rho\) we can implement the arrival model, by performing changes in the following classes:

\begin{enumerate}
    \item In \code{ArrivalFactory} we write a new method\break\code{buildMyModel(parameter1,...)} with any input parameters needed for your model (like a rate-parameter, etc.). In this function we construct the \(\sigma\) and \(\rho\) as symbolic functions. For this we might have to write our own new symbolic functions. These go into the package\break\code{uni.disco.calculator.symbolic\_math.functions}. See the other\break already implemented arrival models for examples.
    \item We add our new model in the list of arrival types in the class \code{ArrivalType}.
    \item To make our new model available in the GUI we need to change the class \code{FlowEditor}
    \begin{enumerate}
        \item First we need to prepare the dialog so it can collect the parameters from users' input. Under the comment-line ``Adds the cards for the arrival'' we can find one card for each already implemented arrival model. We add our arrival model here appropriately.
        \item We add our newly created card to \code{topCardContainer} in the directly subsequent lines.
        \item A bit further down the code we can find the action the dialog should perform after the \code{APPROVE\_OPTION}. We add our own \code{if}-clause and follow the examples of the already implemented flows in how to generate the \code{Arrival}-object from the parameters put in by the user. Make sure to use \code{return;} to jump out of the \code{if}-clause, whenever a parameter could not have been read from the input-fields or was initialized incorrectly (e.g., a negative rate was given).
    \end{enumerate}
\end{enumerate}

\subsection{Adding Service Models}
Adding new service models is completely parallel to how to add new arrival models. Again MGF-bounds must be available to implement a new model (see Definition \ref{def:Service-Bound}). Changes to the code must be made in the classes: \code{ServiceFactory}, \code{ServiceType}, and \code{VertexEditor}. For exact details, compare to the changes being performed for adding new arrival models.
\subsection{Adding a new Analysis}
To add a new method for analyzing a network we follow these steps:
\begin{itemize}
    \item We construct a new class extending the \code{AbstractAnalysis}-class. We must make sure that the output for the \code{analyze}-method produces the required performance measure in an MGF-bound format and is an \code{Arrival}-object.
    \item Next we add the new analysis in the class \code{AnalysisType}.
    \item We add a new case in the class \code{AnalysisFactory}. Should our analysis require more parameters than the one offered by the \code{getAnalyzer}-method of \code{AnalysisFactory} we must make corresponding changes to the factory. When doing so these changes must be propagated to the classes \code{AnalysisDialog} and the \code{analyzeNetwork}-method of the \code{SNC}-class. In this case, however, we would recommend to switch to a builder pattern instead.
\end{itemize}
\subsection{Adding a new Optimization}
To add a new method for optimization a performance bound we follow these steps:
\begin{itemize}
    \item We construct a new class extending the \code{AbstractOptimizer}-class. The new optimizer must define the \code{minimize}-method: The code for how to find a near optimal value goes in here.
    \item Afterwards we add the new optimizer to the \code{OptimizationType}-class and as a new case in the \code{OptimizationFactory}. As with adding new analyses there might be more parameters needed than the optimization factory can currently offer. In this case changes must be propagated to the \code{OptimizationDialog} and to the method called\break\code{optimizeSymbolicFunction} in the \code{SNC}-class. Again, when existing methods have to be changed anyway, it would be advisable to use a more general approach, such as a builder pattern.
\end{itemize}

\section{A Full Example}
\label{sec:full_example}

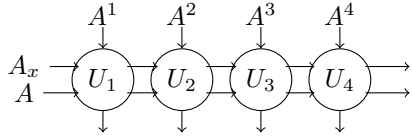
\begin{figure}[t]
	\begin{center}
	\begin{tikzpicture}[scale = 0.7]   
	
	\path (-0.5,0.25) node(Origin_top) {$A_x$}
	++(0,-0.5) node(Origin_bottom) {$A$}
	++(1.5,0.25) node(U_1)[circle, draw]{$U_1$}
	++(1.5,0) node(U_2)[circle, draw]{$U_2$}
	++(1.5,0) node(dots)[circle, draw] {$U_3$}
	++(1.5,0) node(U_n)[circle, draw]{$U_4$}
	++(1.5,0.25) node(dummy_Destination_top){}
	++(0,-0.5) node(dummy_Destination_bottom){};
	
	\begin{scope}[on background layer]
	\path (1,0.25) node(dummy_U_1_top)[text = white]{$U_1$}
	++(0,-0.5) node(dummy_U_1_bottom)[text = white]{$U_1$}
	++(1.5,0.5) node(dummy_U_2_top)[text = white]{$U_2$}
	++(0,-0.5) node(dummy_U_2_bottom)[text = white]{$U_2$}
	++(1.5,0.5) node(dummy_dots_top)[text = white]{$\ldots$}
	++(0,-0.5) node(dummy_dots_bottom)[text = white]{$\ldots$}
	++(1.5,0.5) node(dummy_U_n_top)[text = white]{$U_n$}
	++(0,-0.5) node(dummy_U_n_bottom)[text = white]{$U_n$};
	\end{scope}
	
	\draw[->] (Origin_top) -- (dummy_U_1_top); 
	\draw[->] (Origin_bottom) -- (dummy_U_1_bottom); 
	
	\draw[->] (dummy_U_1_top) -- (dummy_U_2_top);
	\draw[->] (dummy_U_1_bottom) -- (dummy_U_2_bottom);
	
	\draw[->] (dummy_U_2_top) -- (dummy_dots_top);
	\draw[->] (dummy_U_2_bottom) -- (dummy_dots_bottom);
	
	\draw[->] (dummy_dots_top) -- (dummy_U_n_top);
	\draw[->] (dummy_dots_bottom) -- (dummy_U_n_bottom);
	
	\draw[->] (dummy_U_n_top) -- (dummy_Destination_top);
	\draw[->] (dummy_U_n_bottom) -- (dummy_Destination_bottom);
	
	\path (1,1.25) node(Rung1) {$A^1$}
	++(1.5,0) node(Rung2) {$A^2$}
	++(1.5,0) node(Rung3) {$A^3$}
	++(1.5,0) node(Rung4) {$A^4$};
	\draw[->] (1, 1) -- (U_1);
	\draw[->] (2.5, 1) -- (U_2);
	\draw[->] (4, 1) -- (dots);
	\draw[->] (5.5, 1) -- (U_n);
	
	\draw[->] (U_1) -- (1,-1);
	\draw[->] (U_2) -- (2.5,-1);
	\draw[->] (dots) -- (4,-1);
	\draw[->] (U_n) -- (5.5,-1);
			
	\end{tikzpicture}
	\end{center}
	\caption{\label{fig:ladder-topology}A ladder topology for 4 nodes. The flow of interest in this scenario is denoted by $A$. The crossflow $A_x$ takes the same path as $A$, whereas the "rung"-flows $A^i$ contest for resources on one node only.}
\end{figure}
In this section we will give a full walk-through on our modeling steps for the results presented in \cite{Beck:SNCalc2}. In this scenario we consider the topology in Figure \ref{fig:ladder-topology} with 2, 3, or 4 service elements in tandem. In this network we consider the flow of interest as having a low-priority under the crossing flow \(A_x\). This can be interpret as our flow of interest lying in the ``backgronud'' traffic that flows from end-to-end. The rung-flows \(A^1,\ldots, A^4\) are interfering with the service elements in a FIFO- or WFQ-fashion. We also conducted NS3 simulations for the same scenario to make the analytical results comparable.

\subsection{Arrival Model}

We used the following approach for producing arrivals in NS3-simulations: Each arrival consists of a constant stream of data with \(x\) MB/s, where \(x\) is a value that changes each 0.1 seconds and is exponentially distributed. The subsequent values of \(x\) are stochastically independent from each other for all flows and each time-slot. Here the parameters of exponential distributions is chosen, such that the expected datarate for the flow of interest \(A\) is equal to 20 MB/s, the crossflow's expected datarate is 40 MB/s, and each rung-flow's expected rate is\break 20 MB/s.

To model these arrivals in the Calculator we use the MGF-bounds as derived in Example \ref{ex:Exponential-Increments}. Notice that this model is slightly different from the simulations, since the model assumes that the complete bulk of arrivals of one time-slot (with length of 0.1 seconds) arrive in an instant, whereas our simulation streams these arrivals with a constant rate over each single time-slot. We will make up for this difference when modeling service elements.

\subsection{Service Model}

In the NS3-simulation we use a 100 MB/s link-speed between each node. The natural method to model these is to define a constant rate server with rate \(r = 100\) MB/s; however, we want to account for the differences in the model and the simulation when it comes to the flow's burstiness. Note that in the simulations the service elements start working on the data ``as it comes in'', meaning the processing starts from simulation time zero onwards; instead, in our SNC model we would wait one full time-slot and consider all the arrivals of the first 0.1 seconds to arrive in one batch at time \(t= 0.1\) s. As the service rate is constant there is basically a shift of service by one time-slot between the simulation and the model. For this reason we define the service's MGF-bound by the functions
\begin{align*}
    \rho_{S^\prime}(\theta) & = - 10  \\
    \sigma_{S^\prime}(\theta) & = - 10 
\end{align*}
The unit chosen here is MB per time-slot (100 MB/sec = 10 MB/0.1 sec). The above MGF-bound differs from a constant rate MGF-bound by having one additional time-slot of service in \(\sigma_S\), which is available right at the start of the model. We have implemented a corresponding service model into the Calculator as described in the previous Section. 

So far we have not discussed how the service elements schedule the flows. Our simulations work either by a FIFO- or WFQ-scheduling, when it comes to decide whether a packet from the flow of interest or another flow will be processed. So far the Calculator does not have leftover service descriptions implemented for these scheduling disciplines; however, using a leftover scheduling will lead to overly pessimistic results; instead, we have decided to neglect the crossflows' burstiness entirely and subtract the expected number of the rung-flows' arrivals from the constant rate server. This means, we have to subtract the value 2 from our service rates, leading us to the bounding functions \(\rho_S(\theta) = -8\) and \(\sigma_S(\theta) = -8\).

\subsection{End-to-End Analysis}

As a last step we need to account for the crossflow \(A_x\), which joins our flow of interest for the entire path. We need to take this crossflow into account when we want to use Theorem \ref{thm:End-to-End}; in fact, this slightly modifies the proof of this theorem in a straightforward manner. Having this result at hand we implemented a new \code{Analysis} to the Calculator, which uses this end-to-end result. There was no need to implement a new optimization method, as the ones implemented can already cope with this scenario.

With the analysis and the service model implemented the results about end-to-end delay can be produced using the Calculator. Note that calling the analysis repeatedly is needed to produce the graphs in \cite{Beck:SNCalc2}. Since the GUI does not support such a repeated calling we accessed the Calculators methods and classes directly instead. This allowed to automatically loop through an increasing given violation probability.

\bibliographystyle{plain}
\bibliography{references}

\end{document}